# Highly-sensitive detection of the lattice distortion in single bent ZnO nanowires by second-harmonic generation microscopy

Xiaobo Han[1], Kai Wang[1,*], Hua Long[1], Hongbo Hu[1], Jiawei Chen[1], Bing Wang[1,*] & Peixiang Lu[1,2,*]

[1]Wuhan National Laboratory for Optoelectronics and School of Physics, Huazhong University of Science and Technology, Wuhan 430074, China

[2]Laboratory of Optical Information Technology, Wuhan Institute of Technology, Wuhan 430073, China

*Corresponding authors: kale_wong@hust.edu.cn (K. W.), wangbing@hust.edu.cn (B. W.), lupeixiang@hust.edu.cn (P. X. L.)

**Abstract**

Nanogenerators based on ZnO nanowires (NWs) realize the energy conversion at nanoscale, which are ascribed to the piezoelectric property caused by the lattice distortion of the ZnO NWs. The lattice distortion can significantly tune the electronic and optical properties, and requires a sensitive and convenient measurement. However, high-resolution transmission electron microscopy (HRTEM) technique provides a limited sensitivity of 0.01 nm on the variation of the lattice spacing and requires vacuum conditions. Here we demonstrate a highly-sensitive detection of the lattice distortion in single bent ZnO NWs by second-harmonic generation (SHG) microscopy. As the curvature of the single bent ZnO NW increases to 21 mm$^{-1}$ (<4% bending distortion), it shows a significant decrease (~70%) in the SHG intensity ratio between perpendicular and parallel excitation polarization with respect to *c*-axis of ZnO NWs. Importantly, the extraordinary non-axisymmetrical SHG polarimetric patterns are also observed, indicating the twisting distortion around *c*-axis of ZnO NWs. Thus, SHG microscopy provides a sensitive all-optical and non-invasive method for in situ detecting the lattice distortion under various circumstances.

**Keywords**: Lattice distortion; Second-harmonic generation; Polarization-dependent; ZnO nanowires

Nanogenerators can convert nanoscale mechanical energy into electrical power, which is a ground-breaking nanodevice in nanotechnology. Primarily, it is based on the piezoelectric property caused by the bending lattice distortion in ZnO NWs[1,2]. Besides, it is proved that the lattice distortion in nanostructures induced by a controllable mechanical deformation can significantly tune their electronic, magnetic and optical properties[3-8], which shows potential applications in strain sensors[9-11] and field effect transistors (FET)[12]. Many previous experimental and theoretical studies have been devoted to understanding the influence of lattice distortions in nanostructures[13-15]. Therefore, it requires a highly-sensitive method to determine the lattice distortion in nanostructures conveniently. However, HRTEM technique provides a limited sensitivity for distinguishing the tiny variation in lattice spacing (<0.01 nm)[16-18]. Also, it is challenging to study thick samples (>300 nm)[19] and requires high vacuum conditions. X-ray diffraction (XRD) technique is mainly used to study films or powders, while it is not suitable for studying a local-region of samples[20]. Thus, it is still urgent to develop a sensitive all-optical and non-invasive method that is compatible with the body or surface detections under various circumstances.

SHG is an important second-order nonlinear optical process, which converts an electromagnetic wave of frequency $\omega$ into another wave of the doubling frequency $2\omega$[21]. As previously reported, SHG in kinds of nanostructures have been demonstrated to be highly efficient and controllable, showing potential applications in nanoscale light source[22-24] and nonlinear optical microscopy[25]. In particular, the polarization-dependent SHG microscopy is proved to be an alternative method to probe morphology and crystallography due to its high sensitivity on the lattice structures. Recently, it is reported that all three crystal axes orientations of single ZnS NWs can be precisely determined by SHG microscopy[23]. It can be further used to distinguish between wurtzite and zincblende II–VI semiconducting crystal structures and determine their growth orientation[19]. It implies that the polarization-dependent SHG microscopy has the potential in highly-sensitive detection on the lattice variations in nanostructures. However, as far as our knowledge, detections of the lattice distortion in nanostructures by polarization-dependent SHG microscopy have not been addressed yet.

Here we develop a highly-sensitive method to detect the lattice distortion in single bent ZnO NWs based on SHG microscopy. The single ZnO NWs with a *c*-axis growth orientation were bent by a tungsten micro-tip, leading to a bending distortion of the lattice spacing (<4%) confirmed by HRTEM. As curvature of the single bent ZnO NWs increases to 21 mm$^{-1}$, it shows a significant decrease of ~70% in the SHG intensity ratios between perpendicular and parallel excitation polarization with respect to *c*-axis of ZnO NWs. It indicates the one magnitude improvement of the detection sensitivity (~0.001 nm) than that by HRTEM technique. Furthermore, the extraordinary non-axisymmetrical SHG polarimetric patterns are also observed, indicating the twisting distortion around *c*-axis of ZnO NWs. Thus, the SHG-based method provides a sensitive all-optical and non-invasive technique for in situ detecting the lattice distortion under various circumstances.

## Results

**Characterization of the bent ZnO NWs.** ZnO is a typical Ⅱ-Ⅵ semiconductor with a wide band gap of 3.37 eV[15,26], which shows unique properties in potential applications of photoelectric and optical properties[1,3,15,26,27]. Specifically, the wurtzite ZnO belongs to the noncentrosymmetric crystal class of 6*mm*[21], which can generate highly-efficient SHG signals[28-30]. The high quality ZnO NWs were fabricated by chemical vapor deposition (CVD) method, and were further dispersed on the quartz substrate for material characterizations and optical measurements. Fig. 1a,b show the scanning electron microscopy (SEM) images of a single straight and a typical bent ZnO NWs respectively, which reveals the high quality of the as-prepared ZnO NWs with uniform diameters and smooth surfaces. Fig. 1c depicts the low resolution TEM image of the straight ZnO NW. And the corresponding HRTEM image and selected-area electron diffraction (SAED) pattern are shown in

Fig. 1d,e, respectively. The HRTEM image indicates a high-quality single crystal, which shows a growth orientation along [001] (*c*-axis) with a lattice spacing of ~0.26 nm. The observed SAED pattern in Fig. 1e further confirms a highly crystalline structure with *c*-axis aligned along the NW's long-axis (See XRD result in Supplementary Information). Importantly, a single bent ZnO NWs were further analyzed by HRTEM to reveal the bending distortion at the atomic scale. As shown in Fig. 1f, the bending of a single ZnO NW can lead to an inhomogeneous strain contrast in the bent region. Fig. 1g,h show related HRTEM images taken from the tensile and compressive regions, respectively (indicated by white squares in Fig. 1f). It can be seen clearly that the lattice spacing in the tensile region (*c*=0.267 nm) is larger than that in the compressive region (*c*=0.258 nm). Thus, the variation of the bending distortion of lattice spacing in the single ZnO NW is estimated to be 4%, which is in good agreement with the previous works[16-18] (Supplementary Table 1). The variation of lattice spacing could have a strong effect on the second-order nonlinear permittivity of ZnO NWs, leading to remarkable variations of related SHG polarimetric patterns at different curvatures of ZnO NWs.

**Optical SHG measurements in single ZnO NWs.** The SHG properties of the single ZnO NWs were measured by a convenient optical microscopy system, as schematically illustrated in Fig. 2a. The detailed description can be found in the Method section. Fig. 2b shows that a single ZnO NW was bent gradually under the manipulation of a tungsten micro-tip. The curvature of the NWs can be estimated as $K=1/R$, where $R$ is the curvature radius. For theoretical analysis, the relative location of the laboratory frame (XYZ axes) and the crystal frame ($x_c y_c z_c$ axes) was shown in Fig. 2c. The pumping laser propagates along the Y axis, while the *c*-axis ($z_c$ axis) of the single ZnO NW lies along the Z axis. The electric-field of the linearly polarized pumping laser, $E_i$, was rotated on the XZ plane, and angle between $E_i$ and *c*-axis is defined as the polarization angle, $\theta$. And $\varphi$ is defined as the angle between $x_c$ axis and X axis. In the crystal frame, the angle between $x_c$ axis and $y_c$ axis is

120° due to the wurtzite crystal frame. Usually, there is a very small deviation between the substrate and the XZ plane in the Lab frame, indicating a small angle of $\varphi$ in experiment. To avoid the mutual effect between different ZnO NWs, the SHG signals were measured from isolated single ZnO NWs on a quartz substrate. As shown in Fig. 2d, a sharp peak located at ~410 nm is obtained, which is exactly the twice frequency of the pumping laser (~820 nm). In the inset, it clearly shows a quadratic response of the measured SHG intensity as a function of the pumping power, indicating a second-order nonlinear process. Thus, it is concluded that the measured signals can only be ascribed to the SHG responses in a single ZnO NW.

**The SHG polarimetric patterns in a bent ZnO NW.** In order to investigate the polarization-dependent SHG properties as a function of the lattice distortion, the SHG polarimetric patterns were measured at different curvatures of the single bent ZnO NW. A sequential optical image of the same ZnO NW is displayed in Fig. 3a-d with different curvatures of $K$=0, 7.2, 11.1 and 21 mm$^{-1}$, respectively. Noting that the pumping laser spots are fixed at the same position of the ZnO NW (marked by the red dots) during the measurements. Fig. 3e-h present a series of SHG polarimetric patterns as the curvatures of the single ZnO NW increase, corresponding to Fig. 3a-d respectively. The black dots indicate the experimental data, and the red solid curves present the theoretical fits (see Method Section for the detailed procedure). Generally, it can be seen that all the SHG polarimetric patterns are axisymmetrical shown in Fig. 3e-h, which is in good agreement with the previous works[29]. In particular, the SHG intensities at $\theta$=90° indicated by the black arrows dramatically decrease as the curvatures increase, while the SHG intensities at $\theta$=0° are insensitive to the increase of the curvatures.

For a clear evidence, the SHG intensity ratio between $\theta$=90° and $\theta$=0° is plotted at $K$=0, 7.2, 11.1 and 21 mm$^{-1}$ shown in Fig. 4 (red dots). Obviously, the SHG intensity ratio shows a dramatic decrease of ~70% as the curvatures of the single ZnO NW increase to 21 mm$^{-1}$. For a similar curvature, the variation of the lattice spacing in the bent ZnO NWs was estimated to be ~0.01 nm (~ 4 % in relative) by HRTEM technique shown in Fig.1. Since the sensitivity of HRTEM shows a limitation of ~0.01 nm, the SHG-based method presents one magnitude improvement of detection sensitivity (~0.001 nm) than that by HRTEM in consideration of the SHG signal variations in the experiment (See Supplementary Fig.2 and Supplementary Table 1 ). Because the symmetry of the SHG polarimetric patterns keep axisymmetrical, the second-order nonlinear susceptibility tensor for 6*mm* class is also true for the bent ZnO NWs. The theoretical analysis indicates that the SHG intensity at $\theta=0°$ is determined only by component $d_{33}$, while the SHG intensity at $\theta=90°$ is determined only by component $d_{31}$. In Fig. 4, it shows that the fitted absolute value of $d_{33}/d_{31}$ (blue dots) increases as the curvatures of the ZnO NW increase. It implies that the lattice distortion has a significant influence on the component values of second-order susceptibility tensor. Therefore, the lattice distortion can be sensitively detected by measuring the variations of SHG signals.

**The SHG polarimetric patterns in a bent and twist ZnO NW.** Besides the bending distortion, the twisting distortion in the nanostructures is also an important deformation. Especially, it has a great influence on growth of nanomaterials, such as the Eshelby Twist[31,32]. In our experiment, the single ZnO NWs were bending by a tungsten micro-tip. During the bending process, the ZnO NW is possibly twisted around its *c*-axis as illustrated in the inset of Fig. 5a. It can be directly confirmed by SEM technique due to the perfect hexagonal cross-sections of the high quality ZnO NWs. For a clear observation of the facets, the contrast of the SEM images in Fig. 5 was adjusted. Fig. 5c presents the detailed morphology of the middle area of the bent ZnO NW shown in Fig. 5a, the size of both facets aside from the top facet obviously varies along the ZnO NW gradually, indicating a slight twist around its *c*-axis. Fig. 5b,d exhibit the details of both ends of the ZnO NW indicated in Fig. 5a, respectively. The different deflections of the top facets further confirm the twisting distortion in the

NW with a twist angle of ~30°.

In our experiment, the polarization-dependent SHG properties were studied in several tens of NWs. Most of the SHG polarimetric patterns are axisymmetrical, indicating the bending distortion discussed in Fig 3. In particular, the extraordinary non-axisymmetrical SHG polarimetric patterns are also observed in some samples shown in Fig. 6, which indicates the twist distortion during the bending process like in Fig. 5. Fig. 6a-d show the optical images under different curvatures at $K$=0, 11.8, 21.6 and 24.4 mm$^{-1}$, respectively. The red dots indicate the fixed pumping laser spots in the ZnO NW. Fig. 6e-h present a series of SHG polarimetric patterns as the curvatures increase, corresponding to Fig. 6a-d respectively. The measured SHG intensity (black dots) as a function of $\theta$ can be fitted well by red solid curves (see Method Section for the detailed procedure). It presents the extraordinary non-axisymmetric SHG polarimetric patterns with a deflection direction indicated by the blue arrows. And the deflection angle, $\alpha$, is defined clockwise from $\theta$= 90° to the angle pointed by the blue arrows. As the bending curvature increases at $K$=0, 11.8, 21.6 and 24.4 mm$^{-1}$, the deflection angle is estimated to be -16°, 9°, 13° and 24°, respectively. In particular, it shows a nonzero value of $\alpha$ (-16°) at $K$=0 mm$^{-1}$, indicating a twisting distortion even in the straight ZnO NWs. The sign switching of $\alpha$ indicates a reversal process of the twisting distortion around $c$-axis of ZnO NW. It is worth mentioning that the second-order susceptibility tensor of class 6$mm$ is not appropriate for the non-axisymmetric SHG polarimetric patterns during theoretical fitting. Therefore, another crucial nonzero component of the second-order susceptibility tensor, $d_{11}$, was added to the fitting process, which is primarily determined by $\alpha$ (see Method Section for more details). The value of $d_{11}$ is estimated to be 0.7, -0.64, -0.86 and -2.5 pm/V in Fig. 6e-h, respectively. Although the component $d_{11}$ due to the twisting distortion are much smaller than other nonzero ones, it still can be sensitively distinguished by measuring the deflection angle ($\alpha$) in the non-axisymmetric SHG polarimetric patterns. Therefore, the polarization-dependent SHG method can be used in a sensitive detection of the complicated lattice distortion.

## Discussion

With the help of the perfect hexagonal cross-section, the twisting distortion of ZnO NWs is directly observed by SEM technique. However, it requires relatively large diameters of NWs with clear facets and sufficient deflection angles for a clear observation[33]. It would be invalid for the semiconductor NWs with a circular cross-section. Generally, both the analysis of SAED patterns by TEM and the contours observation by bright/dark field TEM[31-33] can be used to calculate the twist in crystals. However, it requires rigorous conditions that a NW must be oriented perfectly and twisted sufficiently for the observation and analysis on the TEM grid[33]. More importantly, it shows a limitation in studying the thick samples (e. g. sub-micrometer), because it requires a sufficient thin sample for ED measurement. In our work, it presents a sensitive and convenient method for detecting the twisting distortion in single ZnO NWs by SHG microscopy. It proved that the SHG-based method has the potentials for detecting the more complicated lattice distortion in nanostructures. Another advantage is that SHG-based method is suitable in various circumstances. Compared with traditional HRTEM and XRD techniques, it can be compatible with different experiment conditions, such as in liquids[25] or in atmospheres, at high or low temperatures. In particular, SHG effect shows a unique tunability in wavelength that can conveniently tune the signal wavelength in a wide region. It shows a higher flexibility in choosing the wavelengths of signals or pumping lasers than other optical methods[34] for specific samples detection. It can be applied deep within crystals by tuning the pumping wavelength, which permits the real-time analysis of the internal lattice distortions and phase transformations in nanostructures. In addition, this method can also be used to detect the lattice distortions in other semiconductor nanostructures, such as nanobelts and nanoplates.

In conclusion, we develop a highly-sensitive method to determine the variation of the lattice distortion in single bent ZnO NWs based on polarization-dependent SHG microscopy. As the bending curvatures of the ZnO NW increase to 21 mm$^{-1}$ (~4% bending distortion), the SHG intensity ratio between $\theta=90°$ and $\theta=0°$ shows a significant decrease of ~70%, and then a high detection sensitivity of ~0.001 nm on the bending distortion is obtained in our experiment. Importantly, the extraordinary non-axisymmetrical SHG polarimetric patterns with a deflection angle $\alpha$ are also observed due to the twisting distortion in the single NWs. The SHG-based technique shows a promising application in detecting the lattice distortion in nanostructures with a high sensitivity. It also provides a convenient all-optical and non-invasive method for in situ detecting, which is compatible with the body or surface detections under various circumstances.

## Methods

**NW Growth and morphology characterization.** ZnO NWs were grown on silicon substrates with a gold catalyst layer under atmospheric pressure in a quartz tube[35]. A mixture of high purity ZnO and graphite powders (1:1 by weight) was placed into a quartz boat as source materials. Flows of 5 sccm oxygen and 100 sccm argon were used as the precursor and carrier gases, respectively. The chamber was then heated to 950°C. After growing for 60 min, the furnace was naturally cooled to room temperature, and the ZnO NWs were collected on the silicon substrates. For further optical measurement, ZnO NWs were transferred onto the quartz substrate by dispersing with ethanol, and the straight ZnO NWs were bent under a microscopy system by a tungsten micro-tip. The morphology and crystal structure of ZnO NWs were visualized through field-emission SEM (FESEM, Nova NanoSEM 450) and TEM (Tecnai G20).

**Optical experiments.** SHG signals of the ZnO NWs were measured under an optical microscopy system, as schematically illustrated in Fig. 2a. To pump the single ZnO NW, a mode-locked

Ti-sapphire femtosecond laser system (Tsunami, Spectra-Physics, ~820 nm, 50 fs and 76 MHz) was focused by a 40× objective (NA=0.55) to a spot diameter of ~4 μm. And the reflected SHG signals were collected by the same objective and directly entrance in a CCD or through a fiber coupled to the spectrometer (Princeton Instruments Acton 2500i with Pixis CCD camera). A 750-nm short-pass filter was placed in front of the CCD and spectrometer to filter out the pumping laser. The combination of a half-wave plate ($A_1$) and a polarizing beam splitter (B) was used to adjust the intensity of the pumping laser. Moreover, another half-wave plate ($A_2$) was used to control the polarization direction of the laser. A tungsten micro-tip fabricated by the electric discharge process was used to bend the single ZnO NWs. All the optical experiments were carried out at room temperature.

**Theoretical calculation of the polarization-dependent SHG intensity**. For theoretical analysis, the electric-field $E_i$ is decomposed into three components $E_{ck}$ ($k=x$, $y$ and $z$) (Supplementary Equations (1)) along the crystal coordinate system, and the SHG polarization components, $P_{ck}$, along the three crystal axes are related to $E_{ck}$ by,

$$\begin{bmatrix} P_{cx} \\ P_{cy} \\ P_{cz} \end{bmatrix} = 2\varepsilon_0 \begin{bmatrix} 0 & 0 & 0 & 0 & d_{15} & 0 \\ 0 & 0 & 0 & d_{15} & 0 & 0 \\ d_{31} & d_{31} & d_{33} & 0 & 0 & 0 \end{bmatrix} \cdot \begin{bmatrix} E_{cx}^2 \\ E_{cy}^2 \\ E_{cz}^2 \\ 2E_{cy}E_{cz} \\ 2E_{cx}E_{cz} \\ 2E_{cx}E_{cz} \end{bmatrix} \quad (1)$$

where the $d_{il}$ matrix is the second-order susceptibility tensor of the ZnO crystal. The values used in the simulation for standard ZnO NWs are $d_{15}$ = 4.2 pm/V, $d_{31}$ = 5.2 pm/V, and $d_{33}$ = −14.6 pm/V.[28] The second-order polarization components in the ZnO NWs can be regarded as electrical dipoles oscillated at the SHG frequency. Considering the collection efficiency, $\eta$, of the objective for an accurate estimation of the polarization-dependent SHG responses, the detected SHG power ($P_{SHG}$) radiated from three dipole antennas is defined as[36],

$$P_{SHG} \propto (\eta_{cx}|P_{cx}|^2 + \eta_{cy}|P_{cy}|^2 + \eta_{cz}|P_{cz}|^2) \quad (2)$$

where $\eta_{ck}$ are the objective collection efficiency for the three SHG polarizations $P_{ck}$. In addition, due to the extraordinary non-axisymmetric SHG polarimetric patterns in bent and twist ZnO NWs, only five non-zero components in second-order nonlinear optical susceptibility tensor for *6mm* class cannot satisfy the demand of fitting. In our experiments, the electric-field component $E_{cx}$ is relatively larger than $E_{cy}$ due to the small value of $\varphi$. It indicates that the polarization component $P_{cx}$ is dominant in determining the angle deflection ($\alpha$) in SHG polarimetric patterns. Therefore, another crucial nonzero component, $d_{11}$, of second-order nonlinear susceptibility tensor (Supplementary Equations (4)) was added during fitting for the twist ZnO NWs. Although the absolute value of $d_{11}$ (<3 pm/V) is smaller than other nonzero components (>4 pm/V), it plays an important role in fitting the non-axisymmetric SHG polarimetric patterns than other zero components in $d_{il}$ matrix.

## Acknowledgment

This work was supported by the 973 Programs under grants 2014CB921301 and National Natural Science Foundation of China (11204097), the Doctoral fund of Ministry of Education of China under Grant No. 20130142110078. We thank Prof. Yihua Gao and Dr. Haixia Li for providing the ZnO NWs. Special thanks to the Analytical and Testing Center of HUST and the Center of Micro-Fabrication and Characterization (CMFC) of WNLO for using their facilities.

## Author contributions

P. X. L. and K. W. conceived the experiments; X. B. H and J. W. C performed the optical experiments；H. B. H, H. L. and B. W. performed theoretical calculations, and K. W. and X. B. H. wrote the manuscript.

# Additional information

**Supplementary Information.** XRD measurement of ZnO NWs; Theoretical analysis of the polarization-dependent SHG response in ZnO NWs; Estimation of the detection sensitivity of the SHG-based method.

# Reference

1. Wang, Z. L. & Song, J. Piezoelectric nanogenerators based on zinc oxide nanowire arrays. *Science* **312**, 242-246 (2006).
2. Yang, R., Qin, Y., Dai, L. & Wang, Z. L. Power generation with laterally packaged piezoelectric fine wires. *Nat. Nanotech*. **4**, 34-39 (2009).
3. Xu, S., Guo, W., Du, S., Loy, M. M. T. & Wang, N. Piezotronic effects on the optical properties of ZnO nanowires. *Nano Lett*. **12**, 5802-5807 (2012).
4. Smith, A. M., Mohs, A. M. & Nie, S. Tuning the optical and electronic properties of colloidal nanocrystals by lattice strain. *Nat. Nanotech*. **4**, 56-63 (2009).
5. Lepetit, T. Optical physics: Magnetic appeal in strained lattice. *Nat. Photon*. **7**, 86-87 (2013).
6. Wu, W. *et al.* Piezoelectricity of single-atomic-layer MoS2 for energy conversion and piezotronics. *Nature* **514**, 470-474 (2014).
7. Rhyee, J.-S. *et al.* Peierls distortion as a route to high thermoelectric performance in $In_4Se_{3-\delta}$ crystals. *Nature* **459**, 965-968 (2009).
8. He, K., Poole, C., Mak, K. F. & Shan, J. Experimental demonstration of continuous electronic structure tuning via strain in atomically thin $MoS_2$. *Nano Lett*. **13**, 2931-2936 (2013).
9. Lipomi, D. J. *et al.* Skin-like pressure and strain sensors based on transparent elastic films of carbon nanotubes. *Nat. Nanotech*. **6**, 788-792 (2011).
10. Yamada, T. *et al.* A stretchable carbon nanotube strain sensor for human-motion detection. *Nat. Nanotech*. **6**, 296-301 (2011).
11. Wu, J. M. *et al.* Ultrahigh sensitive piezotronic strain sensors based on a $ZnSnO_3$ nanowire/microwire. *ACS Nano* **6**, 4369-4374 (2012).
12. Wang, X. *et al.* Piezoelectric field effect transistor and nanoforce sensor based on a single ZnO nanowire. *Nano Lett*. **6**, 2768-2772 (2006).
13. Sun, L., Kim, D. H., Oh, K. H. & Agarwal, R. Strain-induced large exciton energy shifts in buckled CdS nanowires. *Nano Lett*. **13**, 3836-3842 (2013).

14. Cazzanelli, M. *et al.* Second-harmonic generation in silicon waveguides strained by silicon nitride. *Nat. Mater.* **11**, 148-154 (2012).

15. Liao, Z.-M. *et al.* Strain induced exciton fine-structure splitting and shift in bent ZnO microwires. *Sci. Rep.* **2** (2012).

16. Han, X. *et al.* Electronic and Mechanical Coupling in Bent ZnO Nanowires. *Adv. Mater.* **21**, 4937-4941 (2009).

17. Fu, X. *et al.* Outermost tensile strain dominated exciton emission in bending CdSe nanowires. *Sci. China Mater.* **57**, 26-33 (2014).

18. Fu, Q. *et al.* Linear strain-gradient effect on the energy bandgap in bent CdS nanowires. *Nano Res.* **4**, 308-314 (2011).

19. Ren, M., Agarwal, R., Liu, W. & Agarwal, R. Crystallographic Characterization of II-VI Semiconducting Nanostructures via Optical Second Harmonic Generation. *Nano Lett.* **15**, 7341-7346 (2015).

20. Wickham, J. N., Herhold, A. B. & Alivisatos, A. P. Shape Change as an Indicator of Mechanism in the High-Pressure Structural Transformations of CdSe Nanocrystals. *Phys. Rev. Lett.* **84**, 923-926 (2000).

21. Boyd, R. W. *Nonlinear Optics* 3rd edn (Academic Press, 2008).

22. Liu, W., Wang, K., Liu, Z., Shen, G. & Lu, P. Laterally emitted surface second harmonic generation in a single ZnTe nanowire. *Nano Lett.* **13**, 4224-4229 (2013).

23. Hu, H. *et al.* Precise determination of the crystallographic orientations in single ZnS nanowires by second-harmonic generation microscopy. *Nano Lett.* **15**, 3351-3357 (2015).

24. Ren, M. L., Liu, W., Aspetti, C. O., Sun, L. & Agarwal, R. Enhanced second-harmonic generation from metal-integrated semiconductor nanowires via highly confined whispering gallery modes. *Nat. Commun.* **5**, 5432 (2014).

25. Hoover, E. E. & Squier, J. A. Advances in multiphoton microscopy technology. *Nat. Photon.* **7**, 93-101 (2013).

26. Huang, M. H. *et al.* Room-temperature ultraviolet nanowire nanolasers. *Science* **292**, 1897-1899 (2001).

27. Zhu, G., Yang, R., Wang, S. & Wang, Z. L. Flexible high-output nanogenerator based on lateral ZnO nanowire array. *Nano Lett.* **10**, 3151-3155 (2010).

28. Neumann, U., Grunwald, R., Griebner, U., Steinmeyer, G. n. & Seeber, W. Second-harmonic efficiency of ZnO nanolayers. *Appl. Phys. Lett.* **84**, 170 (2004).

29. Johnson, J. C. *et al.* Near-field imaging of nonlinear optical mixing in single zinc oxide nanowires. *Nano Lett*. **2**, 279-283 (2002).

30. Liu, W. *et al.* Near-resonant second-order nonlinear susceptibility in c-axis oriented ZnO nanorods. *Appl. Phys. Lett*. **105**, 071906 (2014).

31. Morin, S. A., Bierman, M. J., Tong, J. & Jin, S. Mechanism and kinetics of spontaneous nanotube growth driven by screw dislocations. *Science* **328**, 476-480 (2010).

32. Zhu, J. *et al.* Formation of chiral branched nanowires by the Eshelby Twist. *Nat. Nanotech*. **3**, 477-481 (2008).

33. Morin, S. A. & Jin, S. Screw dislocation-driven epitaxial solution growth of ZnO nanowires seeded by dislocations in GaN substrates. *Nano Lett*. **10**, 3459-3463 (2010).

34. Chen, J. *et al.* Probing strain in bent semiconductor nanowires with Raman spectroscopy. *Nano Lett*. **10**, 1280-1286 (2010).

35. Li, H. *et al.* Enhanced photo-response properties of a single ZnO microwire photodetector by coupling effect between localized Schottky barriers and piezoelectric potential. *Opt. Express* **23**, 21204-21212 (2015).

36. Jackson, J. D. Classical Electrodynamics 3rd edn ( John Wiley & Sons: Hoboken, NJ, 1999).

## Figure captions

**Figure 1. | Characterization of the ZnO NWs.** (**a-b**) The SEM images of (**a**) a straight ZnO NW and (**b**) a bent ZnO NW on a quartz substrate. (**c**) Low-magnification TEM image of the ZnO NW. (**d-e**) HRTEM image and SAED of the ZnO NW in (**c**), indicating that the nanowire has a high-quality wurtzite structure with the [001] growth direction. (**f**) TEM image of a typically bent ZnO NW. (**g-h**) HRTEM images at the tensile and compressive regions in the bent ZnO NW as marked by the white squares in (**f**), respectively. Both of them show lattice spacing (*c*) of ~0.267 nm and ~0.258 nm, respectively.

**Figure 2. | Optical SHG measurements in bent ZnO NWs.** (**a**) Experimental setup for measuring SHG signal of single ZnO NWs. $A_1$, $A_2$, half-wave plates at 800 nm; B, polarizing beam splitter; $M_1$, an ultrafast mirror with the working wavelength of 700 to 930 nm; $M_2$, a folding mirror. The pumping laser was focused by a 40× objective, and the reflected SHG signal was collected by the same objective. (**b**) sketch of a ZnO NW bent by a tungsten micro-tip. And a geometric model to estimate the local bending curvature ***K*** of the NW, ***K*** = 1/***R***. (**c**) 3D illustration geometry of the lab frame (XYZ) and the crystal frame ($x_cy_cz_c$). The *c*-axis of NWs is along Z-axis. The linearly polarized pumping laser propagates along Y axis. The optical electric field, $E_i$, of the pumping laser is in the XZ plane with a variable ***θ*** to the NW growth axis. Specifically, ***φ*** is the angle between $x_c$ axis and X axis. Inset: the crystal frame in ZnO NWs. (**d**) The spectrum of SHG signal originated from a ZnO NW. Inset: SHG power response as a function of the square of the excitation (average) power ***$P^2$***.

**Figure 3. | Curvature-dependent SHG polarimetric patterns in a single bent ZnO NW.** (**a-d**) Optical microscopy images of the same ZnO NW with curvature of K=0, 7.2, 11.1, 21 mm$^{-1}$, respectively. The fixed position is pumped by laser marked by red dots. Scale bars are 20 μm. (**e-h**) A series of polar plots of the normalized SHG intensity as a function of the polarization angle, ***θ***, at

different curvatures corresponding to the cases in (**a-d**). The experimental data are shown as black dots and the theoretical fits are shown as solid curves. And the arrows indicate the SHG intensity at $\theta = 90°$.

**Figure 4.** | Plot of the SHG intensity ratio between $\theta=90°$ and $\theta=0°$ (red dots) and the absolute value of $d_{33}/d_{31}$ (blue dots) at curvature of $K$=0, 7.2, 11.1, 21 mm$^{-1}$, respectively.

**Figure 5.** | **SEM images of a twist ZnO NW.** (**a**) SEM micrograph of a bent and twist ZnO NW. Scale bar is 10 μm. Inset is a 3D schematic diagram of a bent NW with twisting around $c$-axis. (**c**) SEM micrograph of the typically twist NW within the bent area indicated by the white rectangle in (**a**). Scale bar is 5 μm. (**b**) and (**d**) SEM micrographs of both ends of the NW in (**a**). Scale bars are 1 μm.

**Figure 6.** | **Curvature-dependent SHG polarimetric patterns in a single bent and twist ZnO NW.** (**a-d**) Optical microscopy images of the same ZnO NW at curvature of K=0, 11.8, 21.6, 24.4 mm$^{-1}$ respectively. The red dots represent the spots of the pumping laser, and the position of the spots is fixed. Scale bars are 20 μm. (**e-h**) Polar plots of the SHG intensity as a function of $\theta$ corresponding to the upper optical images. The black dots present the experimental data, and the solid curves indicate the theoretical fits. The blue dashed arrow connects two points of the minimum SHG intensity. (**e**) The deflection angle, $\alpha$, is defined clockwise from $\theta$= 90° to the angle pointed by the blue arrows. The angle $\alpha$ in (**e-h**) is estimated to be -16°, 9°, 13°and 24°, respectively.

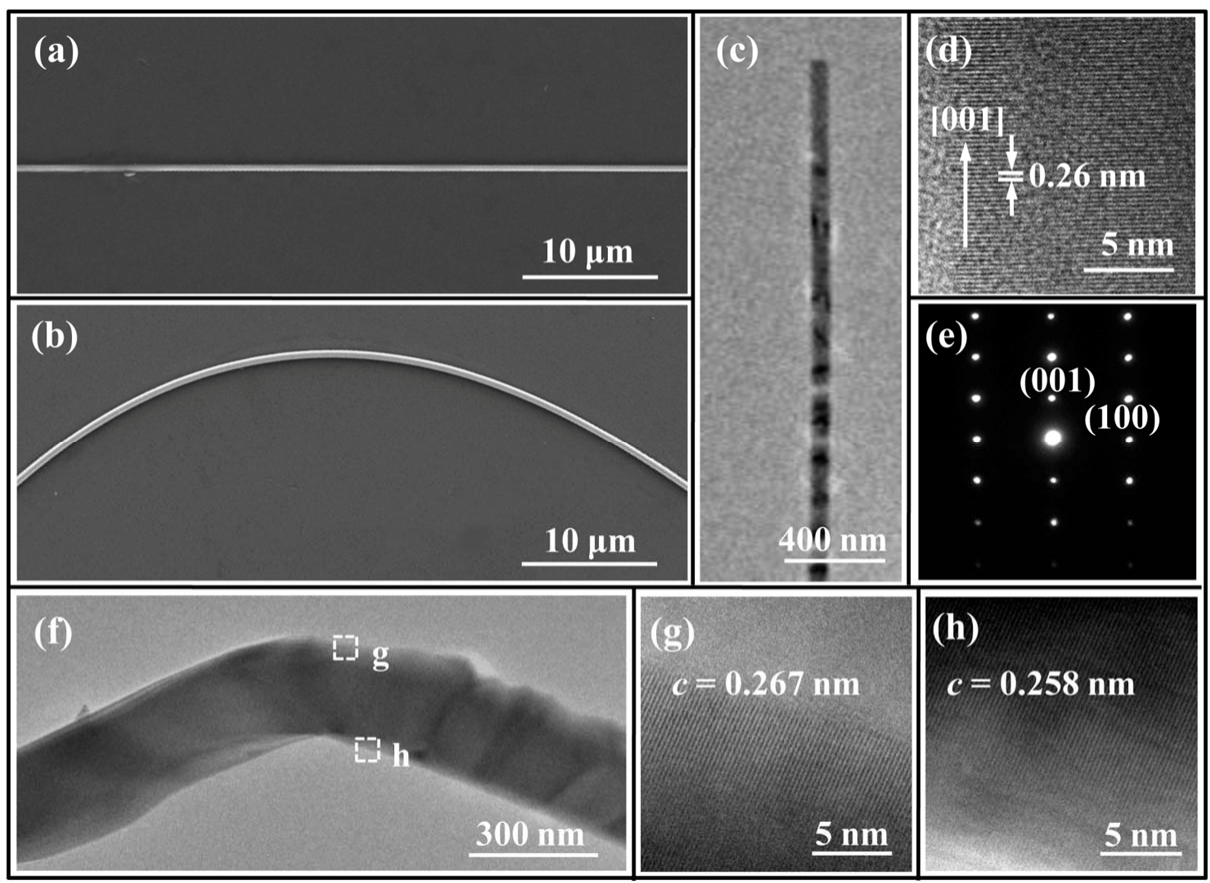

**Figure 1**

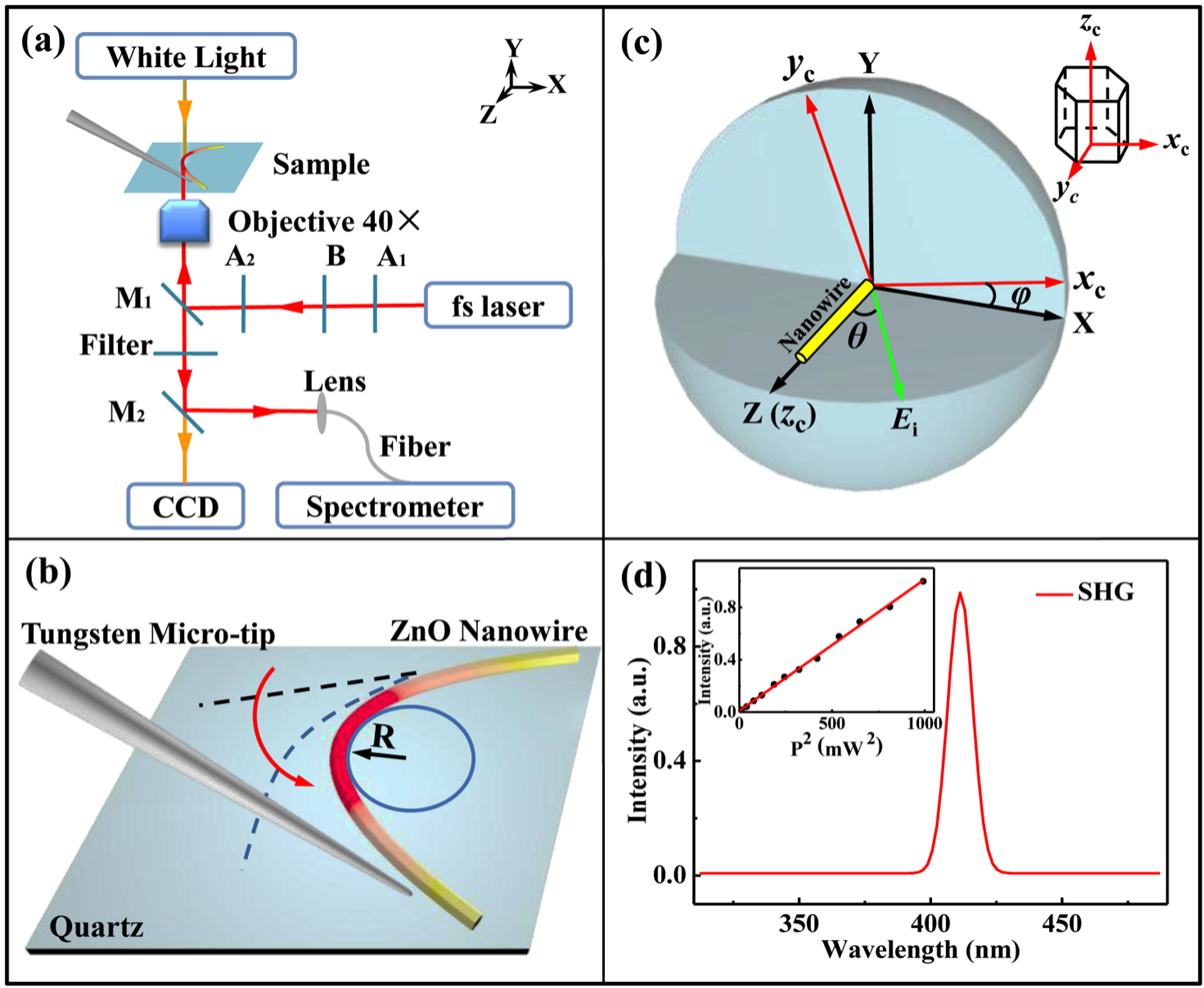

**Figure 2**

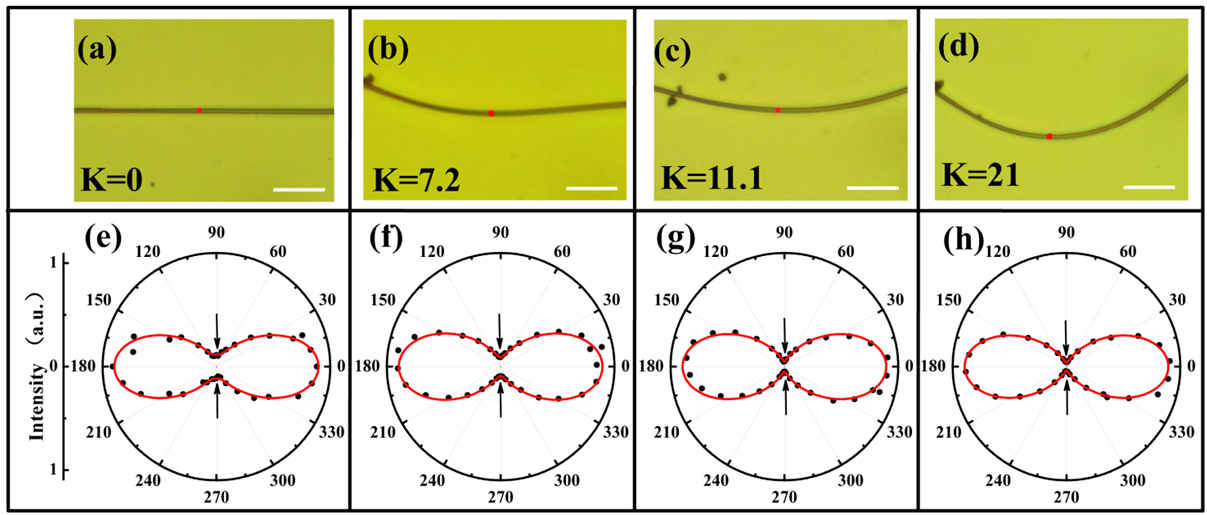

**Figure 3**

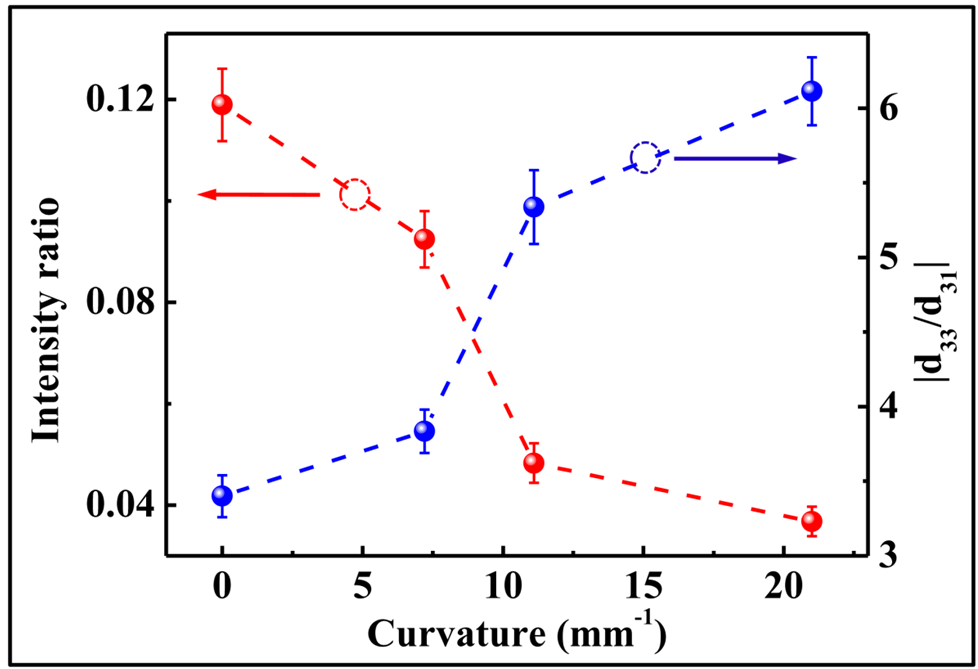

**Figure 4**

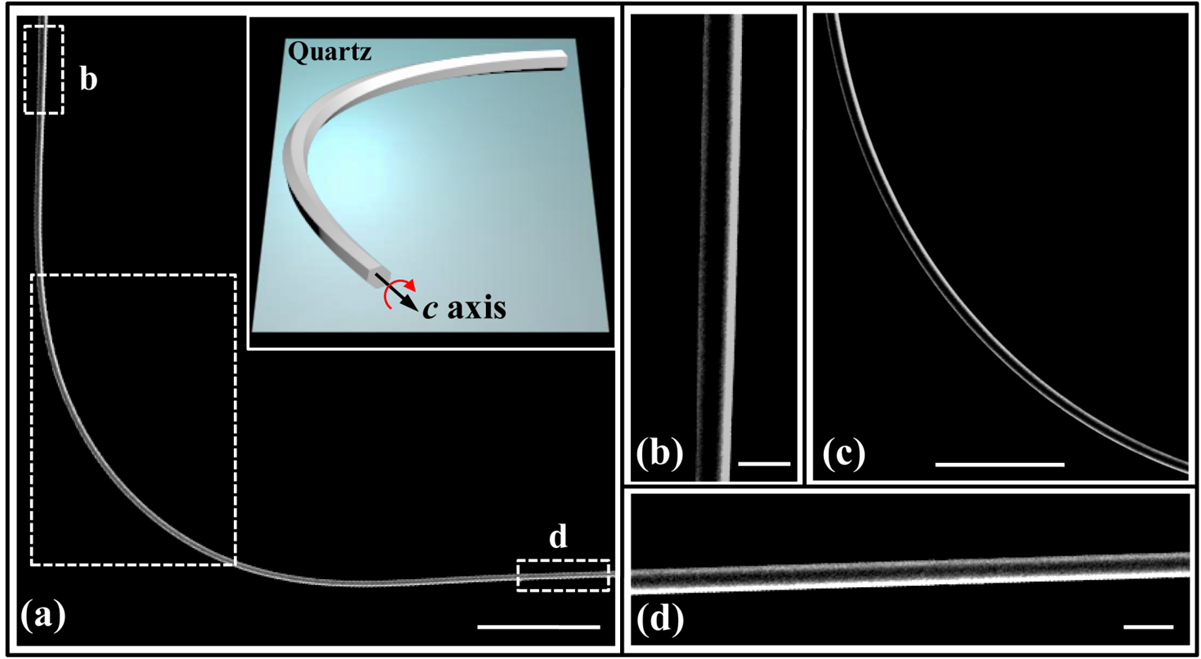

**Figure 5**

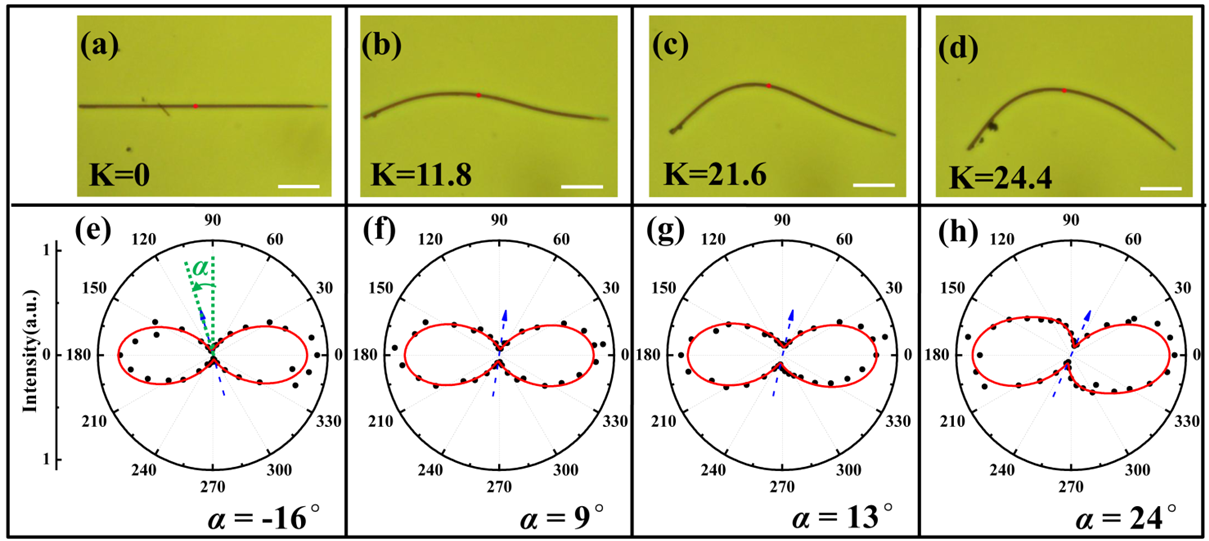

**Figure 6**